# PARA VE MALİYE POLİTİKASININ ARAÇ VE ETKİLERİ: ENFLASYON, KDV VE MEVDUAT FAİZ ORANI İLİŞKİSİ

**Ali DOĞDU* & Murad KAYACAN**

***Öz***

*Ekonomi literatüründe piyasa başarısızlıklarında ve konjonktürel dalgalanmalar sonucu oluşan problemlerde bazı temel araçların uzunca bir süredir kullanılageldiği yadsınamaz bir gerçektir. Ortodoks ve heteredoks iktisadi okulların tercihleri farklı olmakla beraber ağırlıklı olarak tercih edilen Ortodoks yaklaşımda, enflasyonist bir dönemde talebi daraltıcı etki oluşturacak para ve maliye politikaları uygulamada bulunurken, tam tersi olarak deflasyonist bir dönemde ise genişletici etkiler içeren politikalara yönelim artmaktadır. Ülkemizde 1985'ten bu yana alınan Katma Değer Vergisinin de bu nokta da önemi vurgulanmaya değerdir. Öte yandan ekonomik karar alıcı birimlerin piyasada aktör olarak bulunduğu tek nokta maliye politikası değildir. Aynı zamanda bağımsız merkez bankaları ile para politikası araçlarını kullanarak da ekonomiye çeşitli nedenlerden dolayı müdahalede bulunulabilmektedir. İç talebin neden olduğu enflasyonla mücadelede en etkin araçlardan biriside mevduat faizleri olarak görülebilmektedir. Bu çalışmamızda, Türkiye'de 1985-2022 yılları arasında elde edilen KDV gelirleri ile Mevduat Faizi Oranlarının Enflasyon üzerindeki etkisini incelemek amaçlanmıştır. Elde edilen verilerin ekonometrik analizi çerçevesinde ADF birim kök testi, Johansen Eşbütünleşme Testi, Hata Terimleri ve VECM (Vektör Hata Düzeltme Modeli) modelleri kullanılarak analizleri gerçekleştirilmiştir. Analiz sonuçlarına göre, verilerin I(I) düzeyinde durağan olduğu anlaşılmış, uzun dönemde aralarında eş bütünleşik bir ilişkinin var olduğu elde edilmiş ve hata terimi tahmin edilerek VECM analizi çerçevesinde nedensellik bulguları saptanmıştır. Wald Testi nedensellik sonuçlarına göre; Mevduat faizinden KDV ve Enflasyona doğru ve Enflasyondan KDV ve Mevduat faizine doğru (çift yönlü) nedenselliklerine ulaşılırken, KDV'den Enflasyon ve Mevduat faizlerine doğru da bir nedenselliğe ulaşılmıştır.*

* Doktora Öğrencisi, Doğu Akdeniz Üniversitesi, Lisansüstü Eğitim Araştırma Enstitüsü, Ekonomi Anabilim Dalı, ali.dogdu@emu.edu.tr, https://orcid.org/0000-0003-0556-8255

** Doç. Dr., Uluslararası Final Üniversitesi, İ.İ.B.F., Uluslararası Finans Bölümü, murad.kayacan@final.edu.tr, https://orcid.org/0000-0002-7606-6183

# INSTRUMENTS AND EFFECTS OF MONETARY AND FISCAL POLICY: THE RELATIONSHIP BETWEEN INFLATION, VAT, AND DEPOSIT INTEREST RATE

***Abstract***

*It is an undeniable fact that some basic tools have been used for a long time in the economic literature in problems arising from market failures and cyclical fluctuations. Although the preferences of Orthodox and heterodox economic schools are different, in the predominantly preferred Orthodox approach, monetary and fiscal policies that will have a contractionary effect on demand are implemented in an inflationary period, while on the contrary, in a deflationary period, the tendency towards policies with expansionary effects increases. It is worth emphasizing the importance of the Value Added Tax, which has been collected in our country since 1985, at this point. On the other hand, fiscal policy is not the only point where economic decision-making units act as actors in the market. At the same time, it is possible to intervene in the economy for various reasons by using independent central banks and monetary policy tools. Deposit interest can be seen as one of the most effective tools in the fight against inflation caused by domestic demand. In this study, we aimed to examine the effect of VAT revenues and Deposit Interest Rates on Inflation in Turkey between 1985-2022. Within the framework of econometric analysis of the obtained data, the analysis was carried out using ADF unit root test, Johansen Co-Integration Test, Error Terms and VECM (Vector Error Correction Model) models. According to the analysis results, it was understood that the data were stationary at the I(I) level, it was determined that there was a cointegrated relationship between them in the long term, and by estimating the error term, causality findings were determined within the framework of VECM analysis. According to the causality results of the Wald Test; causality is found from Deposit Interest Rate to VAT and Inflation, and from Inflation to VAT and Deposit Interest Rate (bidirectional), while causality is also found from VAT to Inflation and Deposit Interest Rates.*

## Giriş

Ekonomi biliminin araştırma ve faaliyetlerini yoğunlaştırdığı birçok problem ve alan olmakla birlikte; genellikle tam istihdam, fiyat istikrarı, sürdürülebilir büyüme ve cari işlemler açığı gibi odaklanmış olduğu sorunlar dikkat çekmektedir. Bu gibi sorunlar çerçevesinde kamu gelirleri, kamu harcamaları, bütçe süreçleri ve borçlanma gibi ''Maliye Politikası'' çerçevesinde aktif olarak kullanımda tuttuğu araçlar bulunmaktadır. Maliye politikaları sayesinde ekonomide meydana gelen ya da gerçekleşebilecek olan bazı krizlere önlemler alınması sağlanabilmektedir. Hükümetin maliye politikası araçları sayesinde bazı zamanlarda yatırımları teşvik edici, bazı tasarrufları artırıcı ve bazen de tüketim-üretim artırıcı/azaltıcı gibi makroekonomik dengeyi değiştiren ve düzenleyen müdahaleleri olabilmektedir. Bu gibi durumlarda en çok ve yaygın olarak kullanılan araçlardan birisi olarak vergiler gelmektedir. Ülkelerin kendi vergi yasa ve mevzuatları çerçevesinde dolaylı ve dolaysız vergilerden oluşan bir vergi sistemleri mevcuttur. Bu vergi sistemleri arasında en önemli ve toplanılabilmesi nispeten daha kolay olan Katma Değer Vergileri ön plana çıkmaktadır.

KDV çalışmamız çerçevesinde ele alınacak olan enflasyonist bir dönemde oran ve kapsamlarında yapılacak olan değişikliklerle kişi ve kurumların harcanabilirler gelirlerinde meydana getireceği azaltımlar çerçevesinde tüketim üzerinde etkili olabilmektedir. Ayrıca KDV'nin tahsilinin görece daha kolay ve işlevselliğinin daha hızlı tepkiler vermesi bu noktada ülkeler/hükümetler açısından da en çok tercih edilen bir dolaylı vergi çeşidi olmasını sağlamaktadır. Gelişmekte olan ülkelerin birçoğunda bu durum dolaylı vergi gelirlerinin toplam vergi gelirleri içerisindeki payının daha yüksek olmasıyla karşımıza çıkmaktadır. Gelişmiş ülkelerde ise bu oranın daha az olduğu gözlemlenebilmektedir. Diğer taraftan KDV gibi dolaylı vergilerin vergileme adalet ilkesi açısından çeşitli tartışmalara neden olduğu ve aynı zamanda da gelir dağılımında adaletten vergi yüküne birçok farklı mali konuda negatif etkilere sebep olabilmektedir.

Öte yandan piyasada bulunan tüm aktörlerin beklentilerine yön veren, hane halkını ilgilendiren, para miktar, faiz vb. politikalarını içeren ''Para Politikaları'' da aktif olarak kullanılan düzenleyiciler olarak karşımıza çıkmaktadır. Piyasada gerçekleşen herhangi bir kriz ve şok durumunda bu politikaları kullanarak düzenlemeye çalışılmaktadır. İktisadi faaliyetler açısından değerlendirildiğinde yatırım ve tasarruf gibi en önemli etkenleri bir arada etkin bir şekilde kullanıma sunabilmek para politikaları açısından önem arz eden bir konudur. Bazı durumlarda firma ve hane halkı çeşitli krediler vasıtasıyla yatırım yapabilmektedir. Tasarruflar açısından değerlendirildiğinde kişi ve kurumlardan toplanan mevduatların ve bunlara uygulanacak faiz oranlarının piyasanın iş ve işlemleri açından önemli bir etki oluşturacağı yadsınamaz bir gerçektir. Bankaların kaydi para oluşturmalarından en büyük etkiye sahip olan değişkenlerden birisi olarak mevduat faiz oranları karşımıza çıkmaktadır. Bankaların toplamış oldukları mevduatların maliyetini oluşturan bu oran, aynı zamanda bankaların temel gelirlerini oluşturan kredi faiz oranlarına etki etmektedir. Mevduat faizinde meydana gelebilecek artışlar dolaylı olarak kredi faiz oranlarını da artıracağı için, bireysel ve kurumsal kredi satışlarında azalmalar meydana getirebilecek, bu ise tüketimde meydana gelecek azalmaların üretim kanadını etkilemesine, tasarrufların yatırıma dönüşmesine olumsuz etkiler oluşturacaktır. Enflasyonist dönemler ise enflasyonun nedenine bağlı olarak çeşitli ve farklı yönlerde para politikalarının belirlenmesini etkileyerek mevduat faizlerinde değişikliklerin yaşanmasına neden olabilecektir. Mevduat faizlerinin arttığı talep artışına bağlı bir enflasyonist dönemde birey ve kurumların banka tasarruf mevduatlarını artırma eğiliminde olmaları, kredi gibi yeni borçlanma araçlarını kullanmamaları talebi daraltabilecektir.

Bu çerçevede makroekonomik performans açısından uygulanan para ve maliye politikalarının eşanlı ve sürdürülebilir olarak ilerlemesi her ülke için vazgeçilemez bir piyasayı dengeleyici unsur olarak kullanılabilmektedir. Çalışmamızda eşanlı bir şekilde yürütülmesi gereken bu politikaların durumu değerlendirilerek elde edilen sonuçlar çerçevesinde eşanlı olmayan bir para ve maliye politikası uygulamalarındaki sorunlar üzerinde durulacaktır. Aynı zamanda çalışmamızın amacını oluşturan unsur aktif makroekonomik politikalar çerçevesinde maliye politikaları ve para politikalarının enflasyonla mücadelede etkinliği üzerine inceleme yapmaktır. Bu minvalde değerlendirildiğinde, çalışmamızın önemi literatürde henüz değerlendirilmemiş olan Mevduat Faizi, KDV Gelirleri ve Enflasyon üçlüsünde meydana gelen tepkileri güncel bir ekonometrik model çerçevesinde analiz ederek, alanda bulunan bu boşluğa ve soruna spesifik olarak yaklaşarak literatüre katkı sağlamaktır. Literatüre sağlayacağı katkının yanı sıra politika yapıcılar için çalışmamız, enflasyonist bir baskının bulunduğu ekonomide merkez bankası para politikası ve vergi politikalarının oluşumunda etkinliğini artırabilmek adına referans niteliğinde kullanılabilmesi hedeflenmektedir.

## 1. PARA POLİTİKASI VE AMAÇLARI

1776'da Adam Smith yazmış olduğu ‘’Milletlerin Zenginliğinin Doğası ve Nedenleri Üzerine Bir İnceleme’’ kitabında, düzenli bir kağıt para sisteminin ekonomik büyümeyi ve istikrarı sağlamada önemli bir rol oynayabileceği belirtilmiştir (Taylor ve Williams, 2011, s. 830). Para ile ilgili olan etmenlerin neden olduğu Napolyon Savaşları sırasında ortaya çıkan finans krizlerine atıfta bulunarak, 19. yüzyılın başlarında Henry Thornton ve David Ricardo para politikalarının düzenlenmesinin önemini vurgulamışlardır. K. Wicksell, ve I. Fisher I. Dünya savaşı döneminde sırasında hiperenflasyon riskini önlemek için para politikalarının kritik önemini vurgulamışlardır. 20. yüzyılın ortalarında, Büyük Buhran döneminde yapılan hataların tekrarlanmaması için M. Friedman büyüme hızının sabit bir çerçevede gerçekleştirilmesi kuralını önermiş, öte yandan 20.yüzyıl sonlarında, fiyat ve çıktı miktarındaki dengesizlikleri gidermek için kurala dayalı politikalar ön plana çıkmaya başlamıştır. Bu çerçevede Taylor Kuralı gibi çağdaş para politika kuralları önerilerek kullanıma geçmiştir (Doğdu, 2019, s. 29).

Para politikası ve ekonomi politikası birçok açıdan aynı hedeflere sahip olmalarına rağmen, bazı farklılıklar da bulunmaktadır (Doğdu, 2022, s. 451). Benjamin Friedman (2000), para politikasını, genel fiyat seviyesindeki değişimler gibi finansal etkenlerin yanı sıra finans dışı çıktı düzeyi ve iş gücünü içeren iktisadi faaliyetlerin hızı ve yönünü tayin edebilmek için kullanılan bir araç olarak tanımlamıştır. İş gücü, finansal denge, fiyat istikrarı, iktisadi büyüme, gelir dağılımında adalet ve ödemeler dengesinde meydana gelen açık seviyeleri gibi öncelikler para politikalarının öncelikli hedefleri arasında yer almaktadır.

Tam istihdam düzeyi, ekonomide aktif olarak mevcutta bulunan bütün üretim faktörlerinin tam ve etkin olarak kullanılabilmesini ifade eder. Dar bir tanımlamada ise, tam istihdam sadece emeğin etkin olarak ve tam düzeyde kullanılması anlamına gelir. Tam istihdam düzeyine ulaşılmasının para politikaları çerçevesinde amaç olarak belirlenerek işsizliğin ortadan kaldırılması ya da en aza indirgenmesi hedeflenmektedir. Yanı sıra mevsime dayalı ya da arızi olarak gerçekleşen doğal işsizliğe neden olan faktörler etrafında günümüz ekonomilerinde belli bir işsizlik oranının doğal işsizlik seviyesi olarak kabul edildiği bilinmektedir (Doğdu, 2019, s.30).

Merkez bankaları öncelikli ve en önemli hedeflerinden biri olarak fiyat istikrarı tanımlayabilir ve bir ekonomide stagflasyon, deflasyon, slumpflasyon ve enflasyon gibi durumların yaşanmamasını ve fiyatların genel düzeyinin sürekli olarak dengede kalmasını amaçlayabilmektedir (TCMB, 2013, s.8). Fiyatlar belirli bir seviyenin altına veya üstüne çıktığında, bu durum fiyat istikrarını bozar ve yukarıda bahsedilen olumsuz ekonomik durumların ortaya çıkmasına neden olabilir. Enflasyon, fiyatların genel düzeyinde sürekli bir artışı ifade eder, ancak her fiyat artışı enflasyon anlamına gelmez. Fiyat istikrarının sağlanamadığı durumlarda, bireysel karar vericiler ekonomik aktivitelerine riskle yaklaşır, fiyat değişimlerini doğru bir şekilde ayırt edemez ve gereken bilgiye erişemezler (TCMB, 2013, s.10). Fiyatlarda istikrarın gerçekleşmediği zamanlarda, uygulamada olan politikaya güven azabilir ve siyasi iktidar geniş kapsamda iktisadi politikaları uygulayamayabilir. Bu nedenlerle, fiyatlarda istikrarın gerçekleştirilmesi en önemli hedefi olarak nitelendirilebilmektedir (Doğan ve Akbay, 2016, s.630).

Küreselleşen dünyada ekonomik entegrasyonun önemi, serbest piyasa ekonomisi ve dış ticaretteki değişimlerle birlikte daha da artmaktadır. Bu durum, para politikalarının ekonomik yönlendirme sürecinde önemli bir rol oynadığını göstermektedir. Merkez bankalarının ulusal ve uluslararası etkenlere müdahale ederek ekonomiye likidite sağlaması veya liderlik etmesi, para politikalarının ekonomiye etki etme potansiyelini vurgular. Ekonomik gelişme, büyüme ve kalkınma ise dinamik bir iktisat politikasıyla gerçekleştirilebilir. Bu bağlamda, para politikalarının temel amaçlarından birini oluşturur ve ekonomik gelişmenin sağlanmasında önemli bir rol oynar (Doğdu, 2019, 31).

Gelir dağılımı, bir ekonomideki gelir dağılımının adil olup olmadığını ölçmek için Gini katsayısı kullanılarak değerlendirilir. Gini katsayısı 0 ile 1 arasında değer alır; bu katsayısının sıfıra ne kadar yakın olması, gelir dağılımının o kadar adil olduğunu gösterir. Gini katsayısı bire yakın olduğunda ise gelir dağılımında adaletin sağlanamadığını ifade edilmektedir (Tokatlıoğlu ve Selen, 2021: 446). Gelirin yeniden dağıtımıyla adil bir gelir dağılımı sağlanmaya çalışılırken maliye ve para politikalarından faydalanılır. Para politikaları, açık piyasa işlemleri, faiz oranları ve reeskont gibi araçlarla gelirin düzenlenerek adil dağıtılmasına katkıda bulunur. Bireylerin gelirlerini harcama, tasarruf veya yatırım olarak kullanma tercihleri de para politikaları tarafından etkilenebilir (Pehlivan, 2021, s.309). Örneğin, bireyler tasarruf amacıyla fonlarını vadeli bir banka mevduat hesabına yatırır ve faiz oranının sağlayacağı gelir artışlarına ulaşırlar. Bu mevduatlar aynı zamanda başka bireyler veya kurumlar tarafından kredi almak için kullanılabilir ve bu süreçte bankaların kaydi para oluşturma faaliyetleri, Merkez Bankalarının belirlediği sınırlar içinde gerçekleşir.

Faiz istikrarı, ekonomik güvensizliği ve belirsizliği azaltmak için önemlidir. Faiz oranlarında yaşanan dalgalanmalar, ekonomik güveni ve istikrarı zedeleyebilir. Tüketici veya ticari kredi faizlerindeki artışlar, merkez bankası para politikaları çerçevesinde iş gücü, ekonomik büyüme ve fiyat istikrarı hedeflerine aykırıdır ve bu nedenle faiz artışlarını engellemeye çalışırlar. Krediler, para aktarım mekanizmalarına ait bir parça olarak günümüz ekonomilerinde önemli etkilere sahiptir ve finans sistemlerinin, iktisadi büyümenin temel unsuru olarak kabul edilmektedir. Banka kredi hacimlerinde meydana gelen artışlar yatırımları ve tüketim harcamalarında farklı etkiler oluşturabilmektedir etkileyebilir (Çelik ve Kayacan, 2023, s, 1521). Yatırım yapacak firmalar, faiz oranlarındaki belirsizlik nedeniyle gelecek hakkında endişe duyabilir ve yatırımlardan kaçınabilirler. Benzer şekilde, bireysel tüketiciler de örneğin ev alımı gibi büyük kararlarında faiz oranlarındaki belirsizlikten dolayı kredi kullanımını erteleyebilirler. Merkez Bankaları genellikle faiz oranlarının yükselmesini istemezler çünkü bu durum, ellerindeki politika araçlarını azaltarak güçlerini zayıflatabilir. Bu nedenle, faiz istikrarı merkez bankaları için önemli bir konudur (Özpençe ve Noyan, 2023, s.452).

Döviz piyasalarındaki denge, yerel paranın yabancı para cinsinden değerini sürekli olarak korunmasını merkez bankalarının tutum, politikaları ile ilişkilidir. Bu durum, özellikle ulusal paranın değerinin düşmesi durumunda ithalat ve ihracat işlemlerinde sorunlara neden olabilir. Ülke merkez bankaları, piyasada meydana gelen bu istikrarsızlığı düzeltmek için çeşitli araçlar kullanır ve ulusal paranın değer kaybını önlemek için döviz piyasasına müdahale ederler. Ancak bu müdahaleleri yaparken, kendi rezervlerini de dikkate almak zorundadırlar çünkü aşırı rezerv kullanımı Merkez Bankalarını mali açıdan zorlayabilir. Bu nedenle, döviz piyasalarına müdahalede kullanılan araçlardan biri olarak politika faizi işlemleri gerçekleştirilebilir. Bu şekilde sermaye hareketliliğinden faydalanarak döviz sıkıntısını gidermeye çalışırlar. Ancak döviz piyasasındaki sürekli dalgalanmalar güvensizlik ve belirsizlik yaratabilir, bu da iç ve dış yatırımcılar üzerinde olumsuz etkilere neden olabilir. Ulusal paradaki değer düşüklüğünü bertaraf ederek korumak, yüksek oranlı faizlerden kaçınmak merkez bankasının döviz piyasalarının dengede olmasını amaçlar ve bu çerçevede de politikalar belirleyerek kullanır (TCMB, 2013, s.15).

## 2. PARA POLİTİKASININ ARAÇLARI

Para politikası, belirli hedeflere ulaşabilmek için para arzını kontrol ederek piyasaya yön vermek için çeşitli politika araçlarıyla müdahale eder. Bu araçları kullanarak, makroekonomik göstergelerdeki sorunları sürdürülebilir bir şekilde çözmeye çalışır. Aynı zamanda, piyasada oluşabilecek dengesizliklerin ekonomide potansiyel sorunlara yol açmasını engellemek için önlemler alır. Maliye politikasıyla birlikte, piyasada aktif bir rol oynayarak para politikalarının da ekonomide etkin olduğu görülür.

**Tablo.1 Para Politikası Araçları**

| Dolaysız Araçlar | | Dolaylı Araçlar |
|---|---|---|
| Faiz Oranı Kontrolü | Finansal aracı kurumların portföylerini tekrardan düzenlenmesi | Açık Piyasa İşlemleri |
| Reeskont Kotalarının Farklılaştırılması | Özel Mevduatlar | Reeskont Oranları |
| Hisse senetleri ve Tahvil alımları çerçevesinde kredilerin kontrol edilmesi | Tüketici Kredilerinin Kontrolleri | Kısa Vadeli Faiz Oranları<br>Zorunlu Karşılıklar |
| Moral Desteği | Çeşitli Reklam ve Gayri resmi öneriler | Disponibilite<br>Kredi Tavanı |

**Kaynak:** Doğdu, 2019.

### 2.1. Dolaylı Para Politikası Araçları

Merkez bankaları açısından uygulanabilirliği kolay, kullanışlı ve en önemli politika aracı olarak ‘’Açık Piyasa İşlemleri (APİ)’’ gelmektedir. Bu politika aracı piyasada bulunan para arzı ve tabanını kontrol edebilmeyi amaçlarken, hazinenin bonoları ve özel sektörün tahvil ve senetlerini belirlenen sınırlı bir süre karşılığında alıp satabilmesi işlemlerini ifade etmektedir. Doğrudan Satış İşlemleri, Doğrudan Alım İşlemleri, Repo İşlemleri ve Ters Repo İşlemleri ile piyasaya müdahale etmektedir (Parasız, 2010, s.76)

**Tablo.2 Açık Piyasa Alış-Satış-Repo İşlemleri**

| İşlem Türü | Gerçekleştiği Durum | Sonuç |
|---|---|---|
| Kesin Alım | Piyasada uzun süreli likit fazlasının oluştuğu dönem | Piyasada oluşan fazla likiditenin kalıcı olarak çekilmesi |
| Kesin Satım | Piyasada uzun süreli likit sıkışıklığının bulunduğu dönem | Piyasada likidite sıkışıklığının kalıcı olarak giderilmesi |
| Repo | Piyasada geçici süreli likit fazlasının görülmesi | Piyasada fazla likiditenin dönemsel çekilmesi |
| Ters Repo | Piyasada geçici süreliğine likit sıkışıklığının görülmesi | Piyasada likidite sıkışıklığının dönemsel giderilmesi |

**Kaynak:** TCMB, Açık Piyasa İşlemleri Uygulama Talimatı, Ocak 2024.

Öte yandan, merkez bankaları, piyasada oluşan likidite fazlasını gidermek için ticari bankalar nezdinde ihale usulüyle yapmış olduğu nakit alım işlemleri olarak bilinen *mevduat (depo) alım ve satım işlemlerini* gerçekleştirir. Bu işlemler, merkez bankalarının likidite yönetimi aracı olarak kullanılmaktadır. Merkez Bankası Likidite Senet İhraçları, piyasada bulunan fazla likiditeyi azaltmak için kullanılan bir para politikası aracıdır, bu şekilde uygulanan para politikalarının etkinliğini artırmayı amaçlar.

### 2.1.1. Zorunlu Karşılıklar

Banka ve diğer aracı kurumlara aktarılan vadesiz mevduat, özel cari hesaplar, vadeli mevduatlar vb. Merkez Bankası Yasası 44.maddesi gereği, elektronik ödeme aracı çıkarabilen diğer kurumlar ve bankalar, Merkez Bankasında açılan bir hesapta belirlenen oran ve miktarda nakit olarak karşılık bulundurmak zorundadır. Bu karşılık, hesap sahiplerinin mükellefiyetleri esas alınarak belirlenir. Merkez Bankaları, bu oranı değiştirerek, bankaların faiz geliri elde edebileceği mevduat miktarlarını artırabilir veya azaltabilir. (Doğdu, 2019, s. 36).

### 2.1.2. Reeskont Penceresi Politikası

Vadesi gelmemiş olan bir senedin bankalar tarafından vade gününden önce nakde dönüştürülmesi işlemi iskonto olarak adlandırılmaktadır. Bankaların almış olduğu bu senetleri Merkez Bankalarına iskonto ettirmelerine ise reeskont işlemi denilmektedir. Bu işlemlerin gerçekleştirilmesinde merkez bankasının belirlemiş olduğu faiz oranına reeskont faiz oranı denir (Parasız, 2010, s.82).

### 2.1.3. Kısa Vadeli Faiz Oranları

Merkez Bankası, bankalar nezdinde bulunan tahvil, hisse senedi gibi menkul varlıkları bir haftalık vadeyle kabul ederek karşılığında para verdiği ve vade sonunda bu parayı iade ederek menkul kıymetlerini geri aldığı işlem sırasında, uyguladığı faiz oranına kısa vadeli faiz ya da politika faizi oranı denir. Merkez Bankaları, diğer ticari bankalar ve finansal kurumlara uygulamış olduğu politika faizi oranını değiştirerek döviz, hisse senedi, tahvil ve bankalar tarafından verilen kredi miktarları üzerinde etki oluşturabilmektedir. Bu durum merkez bankalarının para piyasalarında istikrar ve kontrolü sağlayabilmek amacıyla kullandığı bir para politikası aracı olarak bilinmektedir (Doğdu, 2019, s. 37).

### 2.1.4. Kredi Tavanı

Bankalar tarafından verilen kredilerde meydana gelen artışlar, toplam talebi de genişleterek talep enflasyonunun oluşmasına, dış borç seviyelerinde meydana gelebilecek artışlara ve dış ticaret açığının büyümesine yol açabilir. Bu tür olumsuz sonuçlardan kaçınmak gelişmekte olan ülkelerde bulunan bankaların oluşturmuş olduğu kredi seviyelerini sınırlamak için kredi tavanı uygulamasına başvurmak gerekebilmektedir. Bu politika rezerv para biriminin ve para stoğunun üzerinde kontrol sağlayabilmekte yardımcı olmaktadır. Merkez Bankalarının uyguladığı daraltıcı para politikaları çerçevesinde, tüketici kredilerini sınırlamak için kredi tavanı belirlenmesi de bu amaçla yapılır (Parasız, 2010, s.84).

### 2.1.5. Disponibilite Politikası

Ticari Bankaların, topladıkları mevduatların önceden belirlenmiş olan yüzdesini kasalarında bulundurmaları zorunludur. Merkez bankları bazı durumlarda bu oranlara müdahale ederek, kredi oluşturma ve kaydi para yaratma süreçlerine sınırlama getirerek ya da tam tersi bir işlemde bulunarak piyasada mevcutta bulunan para miktarını etkileyebilmektedirler (Doğdu, 2019, s.38)

## 2.2. Dolaysız Para Politikası Araçları

### 2.2.1. Döviz Piyasası İşlemleri

Günümüz ekonomilerinde, döviz kuru, dış ticaret, yatırım, tasarruf, enflasyon, büyüme vb. farklı makro değişkenleri Merkez Bankaları etkileyebilmektedirler. Döviz alım satım işlemleri yaparak, piyasada bulunan yerel paranın miktarını ve değerini kontrol edebilir. Merkez Bankaları döviz cinsinden varlıklarını sattığında piyasada bulunan yerel para miktarını azaltırken, tersi işlemde bulunduğunda ise yerel para likiditesini artırır. Döviz kuru artışıyla piyasanın olumsuz etkilenmesini önlemek için piyasaya döviz satışı gerçekleştirebilirler. Tam tersi bir durumda ise, ulusal paranın aşırı değerlendiği dönemlerde bu duruma engel olabilmek için piyasadan döviz alımı gerçekleştirebilirler. Bu işlemleri, döviz kuru ihalesi ile genellikle duyuru yaparak gerçekleştirir ve istikrarı sağlamaya çalışırlar (Parasız, 2010, s.458).

### 2.2.2. Faiz Oranlarının Kontrolü

Merkez Bankaları bazen bankalarda bulunan mevduat ve krediler ile ilgili faiz oranlarını doğrudan kendisi belirleyebilir. Bu faiz oranları, genel faiz seviyelerini etkileyebilir. Örneğin, bankalardaki mevduat faizlerinin düşürülmesi, bankalardaki mevduat miktarının azalmasına ve dolayısıyla Merkez Bankası'nın para arzı politikasındaki etkinliğinin gösterilmesine yol açabilir. Faiz oranlarında bir artış, bireylerin tasarruf etmeyi teşvik ederek tüketimi azaltmasına ve paralarını faize yatırarak gelir elde etmeye yönelmesine neden olabilir. Bu durumda yatırım miktarında bir düşüş görülebilir. Ancak Merkez Bankalarının faiz oranlarına müdahalesi, serbest piyasa ilkeleriyle çatışabilir ve finansal aracılık işlemlerinin daha zor denetlenen piyasalara kaymasına neden olabilir (Doğdu, 2019, s.40).

### 2.2.3. Özel Mevduatlar

Yasal olarak gerçekleştirilen düzenlemeler ile ticari bankalara mecburi olarak aktarılan bazı parasal ödemeler, merkez bankasının para arzını kontrol etmesini sağlayan araçlardır. Örneğin, İngiltere Merkez Bankası geçmişte bazı işlemler yapılırken toplanan mevduatların kendi hesaplarına yatırılmasını zorunluluk haline getirmiş, sonradan ise bu uygulamayı Londra ve İskoçya Takas Bankalarına da uygulamıştır. Bu tür ödemeler arasında ithalat teminatları da yer almaktadır. İthalat teminatları, ithalat işlemini gerçekleştirecek olan firmaların işlem miktarının belirli bir yüzdesini teminat olarak vermelerini içermektedir. Bu teminatlar ithalat işleminin sonuçlanmasına kadar merkez bankasının kasasında kalır ve işlem gerçekleştikten sonra firmaya iade edilir. Teminat oranlarında yapılan artış veya tutulacak olan süreside bir değişiklik para arzını azaltıcı bir etki yaparken, tam tersi bir politika uygulaması para arzını artırıcı bir sonuç doğurabilir. Öte yandan bu politika ithalatı sınırlayıcı bir etki de gösterebilir. Firmalar merkez bankasına vermiş oldukları bu teminat için banka da bulunduğu süre kadar faiz geliri elde edemezler, bu durumda likidite sorunlarının oluşmasına ve ithalat üzerinde negatif bir etkinin oluşmasına sebep olabilmektedir (Doğdu, 2019, s.41).

Bunların haricinde; Finansal Aracıların Portföylerinin Yeniden Düzenlenmesi, Hisse senedi ve Tahvil Alımına Yönelik Kredilerin Kontrolleri, Tüketici Kredilerinin Kontrolleri, Merkez Bankası Moral Desteği, Reklam ve Resmi Olmayan Öğütler gibi farklı araçları da mevcuttur.

### 2.2.4. Ticari Bankalar ve Mevduat Faizleri

Bankalar, ticari bir işletme olarak kabul edilir ve varlıklarını sürdürebilmek ve gelir elde edebilmek için finansal piyasalarda işlemlerde bulunmaktadırlar. Bankalar için fonlar odak noktası olarak ele alındığında, fon arz edenlerin ve fon talep edenlerin arasında gerçekleşecek olan işlemler amacıyla faiz oranlarının temel belirleyici olarak görüldüğü ve bu amaçla bankaların kaynak oluşturabilmenin bir metodu olarak kullanılan faiz oranları, kredi satışlarının gerçekleştirilmesinde bankalar için bir gelir kalemi olarak görülmektedir (Mengüç, 2017, s.2).

Faiz oranları bankaların bilançolarını aktif ve pasif kalemlerini etkilerken bankaların devamlılığı ve karı açısından en önemli etken olarak nitelendirilmektedir. Bankaları diğer özel sektör kurumlarından ayıran temel farklılık, geleceğe yönelik beklentilerin daha fazla önem arz etmesi ile ilişkilendirilmektedir. Bu yüzden, alınacak ya da verilecek olan bir kredinin önemi yüksek ehemmiyet teşkil etmektedir. Alınacak ya da verilecek olan bu kredileri etkileyecek olan faiz oranlarında meydana gelebilecek artış veya azalışlar bankaları finansal riskler açısından daha kırılgan bir hale getirebilecektir (Mengüç, 2017, s.4)

Borçlanmak isteyen kişi veya kurumlar için belirlenecek olan faiz oranları, ekonomik şartlar çerçevesinde belirlenmektedir. Sektörlerin gelişimleri ve finansal açıdan kaynak dağılımları üzerinde faiz oranlarının yapısı ve seviyeleri önem arz etmektedir. Faiz oranı seviyeleri, kaynakların akışında ve yatırım kararı alınmasında rol oynarken, faiz oranlarının yapısı ise risk, likidite ve işlem maliyeti arasında gerçekleşecek olan farkları doğru olarak yansıtmayabilir ve bu durumda da mali açıdan tasarruflarda bir azalma meydana gelebilecektir. Bu yüzden, banka mevduatlarına uygulanacak olan faiz oranları, kredi maliyetlerini etkilerken, kaydi para miktarlarını da etkilemektedir (Mengüç, 2017, s.11).

Bankalar, toplamış oldukları mevduatı kredi talebini karşılama amacıyla kullanırken mevduatlara ödenecek olan faizlerle kredi satışından alınan faizler arasında oluşan fark bankaların gelirlerini oluşturmaktadır (Mengüç, 2017, s.3). Bu yüzden, faiz oranının artırılması ya da mevduat yapısının değişiklikleri kaydi para miktarlarında artış meydana getirebilmektedir. Sonuç olarak faiz oranı hem etken bir nitelikle hem de edilgen bir nitelikle karşımıza çıkmaktadır.

Faiz oranları genel olarak, finansal sistemde faaliyet gösteren ekonomik birimlerin, ülkenin istikrarına ilişkin beklentileri ve devletin borçlanma eğilimi üzerinde etkili olur, bu durumda kredi hacmini etkilemektedir. Aynı şekilde, bankaların faiz maliyetlerini karşılamak için topladıkları mevduatın krediye dönüşmesi (kredi hacminin artması) ülkenin içinde bulunduğu ekonomik duruma bağlı olarak değişmektedir (Vurur ve Özen, 2013, s. 118).

## 3. Maliye Politikaları

Genel olarak devletin mali araçları kullanarak ekonomiye müdahale ederek yön vermesi şeklinde ifade edebilir. Maliye politikalarının konusunu devletin yapacağı kamu harcamaları, borçlanması ve vergiler çerçevesinde mali araçları kullanarak iktisadi olarak hedeflere ulaşılabilmesi için oluşturduğu, uyguladığı ve kullandığı önlemler bütünü olarak değerlendirebiliriz. Maliye politikasını temel iktisadi politikalardan ayrı olarak düşünmek mümkün olmamakla beraber bu politikalarla paralel bir şekilde sürdürülebilir büyüme, tam istihdam, gelir dağılımında adalet ve fiyat istikrarı gibi makro ekonomik sorunların çözümü için kullanılan politikalar bütünü şeklinde görmek daha doğru ve gerçekçi bir yaklaşım olabilmektedir. Maliye politikasının dört temel alt politikasını; Vergi politikası, Kamu Harcamaları (Giderleri) politikası, Borçlanma politikası ve diğer politika ve araçlar oluşturmaktadır (Pehlivan, 2021, s.288).

### 3.1. Kamu Harcamaları (Giderleri) Politikası

Kamu harcamalarının artırılması ve azaltılması ile ekonomide meydana gelecek etkiler etrafında şekillenen bir politika olarak nitelendirilmektedir. Ekonomi içerisinde bulunan aktörlerin harcanabilir gelirlerinin etkileyerek harcama ve toplam talebin düşüş eğiliminde olduğu dönemlerde kamu harcamalarında artırıma gidilerek ekonomik canlanma gerçekleştirilebilir. Çeşitli nedenlerle bu politika çerçevesinde mevcutta bulunan gelirlerine ek olarak fazladan gelir elde eden kişi ve kurumlar bu gelirlerini harcayarak talepte bir artış oluşmasını sağlayarak üretimin canlanmasını ve ekonomik istikrarın oluşmasına katkı sağlamaktadırlar. Öte yandan tam tersi bir durumda ise toplam talepte meydana gelen artış ile karşılaşıldığı durumlarda ise kamu harcamaları azaltılarak talebi düşürücü etki oluşturularak ekonomik açıdan etkin bir politika oluşturulabilmektedir (Pehlivan, 2021, s.75).

### 3.2. Borçlanma Politikası

Ekonomi de talep artışına bağlı olarak meydana gelen baskının iç borçlanma yoluyla birey ve kurumların harcanabilir gelirlerinin azaltılarak müdahale edilmesi, tam tersi bir durumda ise alınan iç borçlanma senetlerinde erken ödeme ile talebi canlandırıcı etkinin oluşturulabildiği bir politikadır (Aksoy, 2011, s.151).

### 3.3. Diğer Politikalar

Temelde bulunan ve yukarıda özetlenen politikalar dışında, dış ticaret politikasından teşvik politikalarına kadar birçok farklı ve çeşitli politikaların tamamını kapsamaktadır.

### 3.4. Vergi Politikası

Vergiler politikası maliye politikası ve genel ekonomik işleyiş açısından son derece önemlidir. Ekonomideki güncel durum ve sorunlar çerçevesinde vergilerin artırılıp-azaltılması şeklinde uygulanan bir politikadır. Artırım ve azaltımlar vergi oranlarında ya da vergi kapsamlarında yapılan değişikliklerden oluşmaktadır. Vergi oranında ya da verginin kapsamında bir artırım olması durumunda kişi ve kurumların harcanabilir gelirlerinde azaltım yapılarak toplam talep üzerinde denetim kurulabilmesi sağlanır. Tam tersi durumda vergi oran ve verginin kapsamında meydana gelen bir azaltım ise kişi ve kurumların kullanılabilir gelirlerinde bir artışa neden olarak toplam talebi yükseltici bir etkide bulunması amaçlanmaktadır (Aksoy, 2011, s.444).

Vergiler, dolaylı vergiler ve dolaysız vergiler olmalarına, konularına veya bireysel ve nesnel olmalarına göre sınıflandırılabilir. Vergilerin adalet ilkelerine uygunluğunu gösteren en önemli ayrım, doğrudan/indirekt vergi ayrımıdır. Vergilerin doğrudan ve dolaylı olarak ayrılmasında dikkate alınan önemli kriterler arasında verginin etkisi, vergi konusunun devamlılığı ve mükellefin önceden belirlenmiş olması bulunmaktadır (Çelik, 2016, s. 256-257).

Mutlak tahsilat süresine sahip, konusu ve mükellefiyeti devamlı olan, bireylerin ve kurumların ödeme gücünün anlaşılmasını amaçlayan ve uygulanması karmaşık olan vergiler; doğrudan vergiler olarak kabul edilir (Cural ve Çevik, 2015, s. 131).

Vergi mükelleflerinin durumları gözetilmeksizin dolaylı vergiler, ekonominin temel unsurlarına, yani üretim, tüketim ve değişime dayanarak alınır. Gelir ve servet elde edildiği anda değil, harcandıkları anda mal ve hizmet fiyatlarına gizlice eklenerek vergilendirilirler. Bu nedenle, vergilendirilmiş malları yoğun kullananlar diğerlerinden daha fazla vergi öderler (Erginay, 1976, s. 121). Bu vergi türü, harcamalar üzerinden alındığı için gelir dağılımını bozma eğilimindedir. Toplanması kolay olduğu ve mali etkisi genellikle fark edilmediği için hükümetler tarafından sıkça tercih edilebilen bir vergi çeşididir. Bu vergi türüne en tipi örnek ise Katma Değer Vergisi olarak verilebilir (Pehlivan, 2021, s.135).

KDV, tüketicilerin ve işletmelerin kullandığı mal ve hizmetler kapsamında ödenen tüketim vergisi olarak gösterilebilir. Her aşamada üretim-tüketim zincirinde alınan harcamalardan kaynaklanır ve ithalattan, mal üretiminden veya satışından elde edilebilir (Omesi ve Nzor, 2015, s. 280). Ancak sonuç olarak tüketiciler tarafından ödenen bir vergidir (Alizadeh ve Motallabi, 2016:337). Gelişmiş ve gelişmekte olan hemen hemen tüm ülkelerde aktif olarak kullanılan KDV, yönetilmesi diğer vergilere kıyasla kolay, vergiden kaçınılmasının ise zor olduğu bir tüketim vergisi olarak kullanılmaktadır (Acharya, 2016, s. 44-55). İyi planlanmış bir KDV, devletin önemli bir gelir kaynağı haline gelir ve politikacılar tarafından genellikle kabul edilir. Ancak, dolaylı vergilerde görülen gerileyici etkiye sahip olması dezavantajıdır. Geniş bir tabana yayılabilmesine karşın tek bir oranda uygulanan KDV, tüketimi kısıtlayıcı bir etki oluşturabilmektedir. Bu nedenle, birçok gelişmekte olan ülkenin vergi sistemlerinde birden fazla oranlı KDV kullanılmaktadır (Smith vd., 2011, s. 3-4).

Katma değer vergisi, bazı yasal durum ve istisnalar harici olmak üzere ekonomideki hemen hemen ayrım gözetmeksizin bütün mal ve hizmet alış-satış işlemlerinden alınan tüketim vergisi olarak karşımıza çıkmaktadır.

Üretim ve tüketim süreçlerinin hemen her aşamasında eklenen değerler üzerinden alınması, yayılı muamele vergilerine benzerliği olarak görülebilmektedir. Bu vergideki katma değer ifadesi, üretim sürecinde bir firmanın başka bir firmalardan aldığı mal ve hizmetlere eklemiş olduğu değeri etmektedir. Kısaca, bir firmanın üretmiş olduğu mal veya hizmetlerin satışlarından elde ettiği brüt gelirleri ile üretimi gerçekleştirebilmek için kullanmış oldukları tüm ara girdilerin maliyetleri arasındaki fark olarak düşünülebilir (Aksoy, 2011, s.375). Bu yüzden, tüm firmaların katkıda bulunduğu değere dayanan bir vergi olan KDV, ekonomideki toplam mal ve hizmet değerini yansıtır ve yerel olarak perakende satış ya da eşdeğer vergisi olarak görülebilmektedir (Alm ve El-Ganainy, 2013, s. 108).

1955'te Fransa'da ortaya çıkan KDV, 1967'de Brezilya'nın kullanımı ile ilk kez Latin Amerika'ya ulaşan ve aynı dönemlerde Danimarka tarafından kabul edilmesi ile Avrupa'ya yayılmaya başlayan dolaylı vergilerin en tipik örneği olarak tarihsel açıdan değerlendirilebilmektedir. Öte yandan Türkiye'de ise 1985 yılında uygulanmaya başlanan Katma Değer Vergisi tek bir oranda alınması durumunda geliri dağılımını bozucu etki oluşturacağından üç farklı oranda uygulanmaya başlanarak bu etkinin azaltılması amaçlanmıştır. Dolaylı bir vergi olarak karşımıza çıkan KDV, tahsilinin kolay olması, tasarruf konusunda artırıcı bir etki oluşturması ve satış fiyatının dışında katma değer üzerinden değerlendirilmesi gibi çeşitli avantajları barındırmaktadır (Şen ve Sağbaş, 2016, s. 217).

Dolaylı vergiler ile enflasyon arasında meydana gelen ilişki incelendiğinde, aralarında çift yönlü nedensellik ilişkisi göze çarpmaktadır. Genel olarak fiyatlar yükselirken, sabit gelire sahip olan bireylerin satın alma gücü azalmaktadır. Bu durum, enflasyonun hızla yükseldiği zamanlarda ise aynı ya da benzer miktarda gelir elde eden bireylerin almış oldukları mal ve hizmet miktarında azalmaya neden olmaktadır. Dolayısıyla, bireylerin tüketim harcamalarındaki azalma, devletin tahsilatını gerçekleştirdiği dolaylı vergi gelirlerinde de bir azalmaya sebep olmaktadır (Aksoy, 2011, s.379). Tam tersi durumda ise; KDV ya da ÖTV gibi dolaylı olan vergi çeşitlerinde oluşan artış, tüketimi azaltıcı bir etki ile sonuçlanacağı bilinen bir kavram olarak görülmektedir. Fiyatlarda artışın olduğu dönemlerde, talep kaynaklı enflasyonu kontrol edebilmek için hükümetler dolaylı vergi oranlarını artırarak, enflasyonist baskıyı kırıcı bir araç olarak kullanabilmektedirler (Mutlu ve Çelen, 2012, s. 75). Öte yandan özellikle dolaylı vergilerin oranlarında meydana gelecek artış üretici kanadında olumsuz etki oluşturarak maliyet enflasyonu oluşturacağı bilinmektedir. Devletin uyguladığı vergi tarifeleri üretim maliyetlerinde artışa neden olarak toplam arzda azalmaya ve enflasyonun temel sebeplerinden birisi olarak kabul edilmektedir (Yurttagüler ve Horvath, 2022, s. 130).

Maliye politikalarının temel amaçlarının başında ekonominin istikrarı gelmektedir. Ekonomide oluşabilecek bir sorun ya da piyasa başarısızlığına maliye politikaları ile müdahalede bulunarak sürdürülebilirlik çerçevesinde ekonomik istikrarın tekrar tesisi sağlanmak amaçlanmaktadır. Bu çerçevede ekonomide deflasyon ve enflasyon gibi farklı dönem ve evreler oluşabilmektedir (Pehlivan, 2021, s.288). Deflasyonist bir baskının oluştuğu noktalarda ekonomik müdahalenin para ve maliye politikaları çerçevesinde paralel bir şekilde yürütülerek gerekli işlem ve araçlarla çözümlenmesi sağlanmaktadır (Pehlivan, 2021, s.289). Öte yandan bu çalışmanın da temel değişkenlerinden bir olan enflasyonist dönemlerde ise; maliye politikalarının işlevi ve kullanılabilir olmasının yanı sıra para politikaları ile tam uyum içerisinde hareket etmesi gerekmektedir (Pehlivan, 2021, s.292). Aynı zamanda enflasyonist dönemlerde maliye politikalarının etkin olabilmesi enflasyonun nedenine bağlı olarak kullanılacak politika ve araçların seçilmesine bağlıdır.

## 4. Enflasyon ve Maliye Politikası

Fiyatlar genel seviyesindeki artışları ifade eden bir olgudur. Enflasyon kendi içinde nedenlerine göre üçe ayrılmaktadır. Talep enflasyonu; ekonominin tam kapasite de olduğu durumlarda farklı nedenlere dayalı olarak toplam talepte oluşan artışın üretim kapasitesini genişleterek karşılanamadığı durumlarda fiyatlar genel seviyesinde meydana gelen yükselme olarak tanımlayabiliriz. Maliyet enflasyonu; üretim girdi fiyatlarında meydana gelen artışlar sonucunda maliyetlerde gerçekleşen artışların fiyatlara yansıması şeklinde nitelendirilebilmektedir (Pehlivan, 2021, s.296). Türkiye gibi gelişmekte olan ve enerji açısından dışa bağımlı olan bir ülkede maliyet enflasyonunu oluşturan temel etmenlerden biri olarak enerji kalemi verilebilir. Enerji kavramının sadece bizim için değil aynı zamanda gelişmiş ekonomilere sahip ülkeler içinde önemli bir girdi kalemi olduğu yadsınamaz bir gerçektir. Örneğin, literatüre petrol krizi olarak girmiş olan, 1973’te enerji maliyetlerindeki yüksek artışlar global bir krizi neden olmuştur (Doğdu, 2022, s. 59). Son olarak ise ele alabileceğimiz enflasyon türü, ekonomide bulunan yüksek talebe karşın, atıl kapasitenin bulunduğu bir ekonomide yapısal sebeplerden dolayı üretimde gerçekleşmeyen artış fiyatları yükseltebilmektedir. Bu durumda ise oluşan enflasyon yapısal bir enflasyon olarak karşımıza çıkmaktadır (Eğilmez, 2024).

Enflasyon gelir dağılımından faizlerdeki yüksek oranlara, döviz kurlarındaki artışlardan üretimin düşmesine kadar ekonomide birçok farklı etkiye neden olabilmektedir. Öte yandan enflasyon ekonomi dışında sosyal ve siyasal olarak da çeşitli negatif etkilere yol açabilmektedir. Ulusal para biriminde meydana gelen ani ve aşırı düşüşler ekonomik aktörlerin tasarruf, harcama vb. birçok farklı alışkanlığını değiştirmekte ve bununda sosyolojik ve politik sonuçları olabilmektedir. Enflasyonla mücadele bağlamında, her ülkenin Merkez Bankası'nın etkisi, kendi gelişmişlik düzeyine göre farklılık gösterebilir. Genel olarak, hemen her ülkede iktisadi açıdan bir merkez bankası, parasal ve finansal tüm politikaları düzenleme yetisine sahiptir.

Ancak, bu araçların etkinliği, döneme, hedeflere ve ülkenin gelişmişlik seviyesine bağlı olarak değişiklik gösterebilir (Doğdu, 2023, s. 168).

## 5. Ampirik Literatür

Tablo 1'de sunulan bilgiler, konuya ilişkin gerçekleştirilen bazı çalışmaların ekonometrik özetini içermektedir. Tabloya bakarak konuyla ilgili yapılan araştırmalar hakkında genel bir bakış elde edilebilir.

**Tablo 3. Literatürde Yapılan Ampirik Çalışmalar**

| Yazar | Değişken | Yöntem | Sonuç |
|---|---|---|---|
| Nar, 2020 | Enflasyon, Cari Denge, Bireysel Kredi Kartları, Bireysel Konut Kredileri, Bireysel Taşıt Kredileri, Bireysel Tüketici Kredileri | VAR Modeli, Granger Nedensellik | TÜFE tarafından taşıt kredilerine ve kredi kartlarına doğru tek yönlü bir nedensellik ilişkisi gözlemlenirken, TÜFE'den (enflasyondan) kaynaklanan tek yönlü nedensellik ilişkisi konut kredileri, kredi kartları ve taşıt kredilerinden cari dengeye doğrudur. Kredi kartlarından konut kredilerine doğru tek yönlü bir ilişki bulunurken, taşıt kredileri ile kredi kartları arasında çift yönlü bir ilişki mevcuttur. |
| Kurt, 2022 | Dolaylı-Dolaysız Vergiler, Enflasyon, Büyüme | VAR Modeli, Granger Nedensellik | Dolaysız vergilerden enflasyona doğru tek yönlü nedensellik belirlenirken, dolaysız vergiler ile ekonomik büyüme arasında doğru yönlü ve dolaylı vergiler ile ekonomik büyüme açısından ise çift yönlü nedensellik bulunduğu tespit edilmiştir. |
| Yurttagüler ve Kutlu Horvath, 2022 | Dolaylı vergi gelirleri ile Tüketici fiyat endeksi | Johansen Eşb., Granger Nedensellik | Vergi gelirleri ile TÜFE değişkenleri arasında tek yönlü bir nedensellik ilişkisi belirlenirken, TÜFE değişkeninin vergi gelirlerinin sebebini oluşturduğu, ancak vergi gelirlerinin TÜFE'nin Granger nedeni olmadığı sonucuna varılmıştır. |
| Hazman ve Bıçaksız, 2022 | 2005:01-2021:11 aylık veriler ile dolaylı vergiler ve enflasyon | VAR Modeli, Granger Nedensellik | Analiz sonuçlarına göre seçili dönemde enflasyon ile dolaylı vergiler arasında çift yönlü nedensellik olduğunu aktarmışlardır. |
| Kaya ve Öz, 2016 | Enflasyon oranı (TÜFE), para arzı, bütçe açığı ve reel döviz kuru | ARDL-ECM | Analiz sonuçları çerçevesinde, enflasyon ve para arzı arasında uzun dönemde, anlamlı ve pozitif bir ilişki bulunmuş, enflasyon ile bütçe açığı arasında ise istatistiksel olarka anlamlı bir ilişki bulunamamıştır. |
| Akdoğan, 2020 | Türkiye‟de 2004-2019 dönemi aylık verileri, enflasyon oranları (ÜFE-TÜFE) ile dolaylı vergi gelirleri | VAR-Granger-Toda Yamamoto-ARDL | Analiz, dolaylı vergi gelirleri ile üretici fiyat endeksi (ÜFE) arasında nedensellik ilişkisi bulunmazken, tüketici fiyat endeksi (TÜFE) ile dolaylı vergi gelirleri arasında ise tek yönlü bir nedensellik ilişkisi olduğu aktarılmıştır. Öte yandan, TÜFE ile dolaylı vergi gelirleri arasında uzun dönemde ve negatif bir ilişki olduğu tespit edilmiştir. Bu durum, ilgili dönem itibari ile enflasyon oranındaki artışın dolaylı vergi gelirlerinin reel değerini azalttığını ifade etmişlerdir. |
| Koçak, Karış ve Çil, 2022 | Türkiye için 1965-2019 yıllık Enflasyon-Enflasyon Oranı-Vergi Gelirleri | Enflasyon oynaklığı EGARCH-ADF-PP-VAR Modeli | Sonuçlara göre vergi gelirlerinin enflasyon oynaklığından büyük ölçüde etkilendiğini göstermektedir. |
| Akçacı ve Koçağ, 2013 | 2006:01-2012:12 aylık veriler personel, transfer, mal-hizmet harcamaları, faiz ödemeleri ve enflasyon | Johansen Eşbütünleşme-Granger Nedensellik-VECM-VAR | İlgili değişkenler arasında Pers. Trs. Mal ve Hizm. ve faiz ödemelerinden enf. Doğru kısa dönemde tek yönlü nedensellik, kamu serm. Harcamalarından enf. Doğru nedensellik bulunmamıştır. Uzun dönemde ise, TÜFE'den kamu harcamalarına, kamu harcamalarından enflasyona doğru nedenselliğe ulaşılmıştır. |
| Arısoy ve Ünlükaplan, 2011 | Aylık veriler 1994:1-2010:6 KDV Oranı-Kamu Harcamaları-KDV Gelirleri-TEFE | Johansen Eşbütünleşme-VECM Etki Tepki-Varyans Ayrıştırması | Etki-Tepki Analiz sonuçları, Varyans Ayrıştırması paralelinde KDV gelirleri ile Kamu Harcamaları arasında karşılıklı ve etkili bir ilişki olduğunu, enflasyonist etkiler taşıdığını göstermektedir. |
| Demir ve Balkı, 2023 | Türkiye genelinde online ve yüz yüze anket tekniği kullanılarak 1.100 kişi | OLS regresyon analizi | Bireylerin genel olarak katma değer vergisi indirimlerine yönelik algısı olumlu yöndedir. Bu, piyasanın canlanmasına yönelik uygulanan katma değer vergisi indirimi politikasına katkı sağlayabilir. Ayrıca, bireyler, katma değer vergisi indirimlerinin fiyatlara yansımadığını düşünmektedir. 2022 yılı Ocak-Ekim döneminde tüketici fiyat endeksinin sürekli artması, bireylerin algısıyla örtüşmektedir ve indirimlerin fiyatlara yansımadığı düşüncesini desteklemektedir. Dolayısıyla, katma değer vergisi indirimleri politikasının fiyatları baskılamak ve düşürmek açısından yeterli olmadığına dair bir görüş ifade edilebilir. |
| Akıncı ve Özçelik, 2018 | Türkiye'de 2006:M1-2018:M5 döneminde dolaylı vergiler ile enflasyon | ARDL | Dolaylı vergiler ile enflasyon arasında uzun dönemli bir eşbütünleşme ilişkisi bulunduğu belirlenmiştir. |
| Akgül, 2022 | 2010M1-2021M6 dönemini kapsayan aylık | Hatemi-J (2012) ve Hacker ve Hatemi-J (2006) simetrik nedensellik analizi | Analiz sonuçlarına göre enflasyondan vergi gelirlerine doğru tek yönlü simetrik bir nedensellik bulmuştur. Enflasyon negatif şokundan vergi gelirleri pozitif şokuna doğru bir asimetrik nedensellik tespit edilmiştir. Öte yandan asimetrik etki-tepki fonksiyonu sonuçlarına göre enflasyon negatif şoklarının zamanla vergi gelirlerini artırdığını gözlemlenmiştir. Bu durum, Türkiye'de Tanzi etkisinin güçlü olduğu vurgulamışlardır. |
| Yaraşır, Tülümce ve Yavuz, 2019 | KDV-GSYH 1985-2018 | Lee ve Strazicich tek kırılmalı birim kök testi-Gregory-Hansen tek kırılmalı eşbütünleşme testi-FMOLS | KDV, vergi sistemimizde önemli bir yer tutan gelir yaratıcı bir vergidir. Araştırmanın sonuçlarına göre, KDV gelirlerindeki %1'lik bir değişim, toplam vergi gelirlerini 2,84 birimlik artışa neden olmaktadır. Öte yandan, ekonomik büyüme ile KDV arasında pozitif bir ilişki olduğu bulunmuştur. KDV gelirlerindeki %1'lik bir değişim, GSYH'de 13,45 birimlik bir artışla ilişkilendirilmektedir. |
| Çatalok, 2019 | Enflasyon Oranı ve Toplam Vergi Gelirleri | EKK Yöntemi | Araştırma sonuçları, enflasyonun vergi gelirleri üzerindeki olumsuz etkisinin olduğunu göstermektedir. Vergi gelirleri nominal olarak her yıl artmasına rağmen, enflasyon nedeniyle reel olarak azaldığını ve bu durumun vergilerin hesaplanması ve tahsilatı arasındaki zaman farkından kaynaklandığını aktarmıştır. |
| Çiftci, 2015 | 1980-2014 dönemine ait yıllık TÜFE ve GSYİH | VAR-ADF-Granger Nedensellik Analizi | Enflasyondan büyümeye doğru nedensellik ilişkisi olduğu belirlenmiştir. Yapılan regresyon denkleminde enflasyon değerinin -0,6 olduğu bulunmuştur. Kısaca, %1 birimlik enflasyon artışı, büyümeyi %0,6 oranında azaltmaktadır. |

Konu çerçevesinde literatür incelendiğinde birçok farklı araştırmacının bu konu üzerinde yoğunlaştığı ve çeşitli değişkenlerle farklı analizleri kullanarak istatistiksel değerlendirmeler yaptığını görebilmekteyiz. Yapılan çalışmaların enflasyon, dolaylı-dolaysız vergiler, GSYH, bütçe gelir ve gider kalemleri etrafında şekillendiğini, bu çalışmalarda da ekonometrik olarak farklı model ve analizlerin kullanıldığı anlaşılmaktadır.

## 6. Veri Seti ve Metodoloji

Bu çalışmamızda, Türkiye'de 1985-2022 yılları arasında elde edilen KDV gelirleri ile Mevduat Faizi Oranlarının Enflasyon üzerindeki etkisi incelenmektedir. Bu sebeple Merkez Bankası EVDS sisteminden ve Hazine ve Maliye Bakanlığı'ndan analizde kullanılacak olan veriler elde edilmiştir. Elde edilen verilerin ekonometrik analizi çerçevesinde ADF birim kök testi, Johansen Eş Bütünleşme Testi, Hata Terimleri ve VECM (Vektör Hata Düzeltme Modeli) modelleri kullanılarak analizleri gerçekleştirilmiştir.

**Tablo 4. Çalışmada Kullanılan Değişkenler**

| Değişken | Açıklama | Kaynak |
|---|---|---|
| KDV | 1985-2022 Katma Değer Vergisi Gelirleri | Hazine ve Maliye Bakanlığı |
| Mevduat Faizi Oranları | 1985-2022 Mevduat Faizi Oranları | Merkez Bankası EVDS |
| Enflasyon | 1985-2022 Enflasyon Oranları | Merkez Bankası EVDS |

## 6.1. Durağanlık Testleri

Çalışmamızda veri setimizin Johansen Eşbütünleşme ve diğer yöntemleri kullanabilmemiz için verilerin durağanlıkları kontrol edilmiştir. Bu kapsamda değişkenlerin literatürde en çok kullanılan yöntem olan ADF kapsamında durağanlık seviyeleri analiz edilmiştir. ADF testi, standart DF (Dickey-Fuller) testinin geliştirilmiş bir versiyonudur ve AR(p) sürecinden faydalanır. ADF testi, p gecikmeli fark terimlerini denklem sistemlerine ekleyerek seriler arasındaki yüksek dereceli korelasyon ilişkisini incelemektedir. Uygun gecikme sayılarını belirlemek için AIC (Akaike Information Criterion) ve SIC (Schwarz Information Criterion) kriterleri kullanılır (Bozkurt, 2013, s. 41).

ADF testinde, $H0$ hipotezi ilgili veride birim kök bulunduğunu, kısaca durağanlığa sahip olmadığını, $H1$ hipotezi ise serinin durağan olduğunu belirtir. Bu nedenle, hesaplama sonuçlarına göre test istatistiğinin kritik değerden büyük olduğu durumlarda, $H0$ hipotezi reddedilir ve serinin durağan olduğu sonucuna varılır. Analiz sonuçlarına göre çalışmada yer alan tüm değişkenlerin Sabit ve Trendli birinci seviye farklarda durağan oldukları sonucu elde edilerek diğer analiz yöntemlerine ilerleme kat edilmiştir.

**Tablo 5. Durağanlık Testi Sonuçları**

| Değişkenler | Değişkenin Durumu | Sabit Trendsiz | Sabit ve Trendli |
|---|---|---|---|
| KDV | Seviye | -1.532032 | -0.491650 |
| | Birinci Fark | -7.143765* | -7.214564* |
| Mevduat Faizi Oranları | Seviye | -1.160549 | -2.205947 |
| | Birinci Fark | -5.790699* | -5.734187* |
| Enflasyon | Seviye | -1.170389 | -0.796213 |
| | Birinci Fark | -4.256429 | -4.186534* |

## 6.2. Johansen Eşbütünleşme Testi

Çalışmada kullanılan değişkenlerin birinci farklarının durağan olduğu göz önüne alındığında, ilgili verilerde Johansen eşbütünleşme testi uygulanabilecektir. Johansen eşbütünleşme testi, değişkenlerin arasında bulunan uzun dönemli bir ilişkinin varlığını belirtmekte ve temelde VAR analizine göre uygulanmaktadır. Bu yüzden, eşbütünleşme testi yapılmadan önce optimal gecikme uzunluğu belirlenmelidir. Değişkenlerin optimal gecikme uzunluğu 1 (k=1) olarak elde edilmiştir. Bu nedenle, birinci dereceden durağan (d=1) serilerin uygun gecikme uzunluğu 1 (k=1) modeli, Johansen eşbütünleşme testi için Eviews 12 programı kullanılarak gerçekleştirilmiştir.

**Tablo 6. Gecikme Uzunluğunun Belirlenmesi**

| Gecikme | LR | FPE | AIC | SC | HQ |
|---|---|---|---|---|---|
| 0 | NA | 876.0381 | 15.28899 | 15.42503 | 15.33476 |
| 1 | 38.60453* | 400.6993* | 14.50325* | 15.04743* | 14.68635* |
| 2 | 11.59733 | 449.7279 | 14.60265 | 15.55498 | 14.92308 |
| 3 | 12.32028 | 472.3627 | 14.61244 | 15.97290 | 15.07020 |

**Tablo 7. Destekleyici İstatistik Sonuçları**

| Normallik Testi Sonuçları | | LM Testi Sonuçları | |
|---|---|---|---|
| Mean | -1.23e-16 | F-İstatistiği | 124 |
| Median | 0.159533 | R Kare | 32,11365 |
| Maximum | 2.231.974 | Ki Kare | 0.000 |
| Minimum | -2.443.093 | Olasılık | 0.000 |
| Std. Hata | 1.463.397 | Seride oto-korelasyon sorunu yoktur. | |
| Skewness | -0.104186 | | |
| Kurtosis | 1.732.370 | Koentegrasyon Test Model Seçim Sonuçları | |
| Jarque-Bera | 2.475.456 | İz Değeri | Max. Öz Değer |
| Probability | 0.290042 | 1 | 1 |
| Seri normal dağılıma sahiptir. | | | |

Johansen-Juselius (1990) tarafından açıklandığı gibi, her serinin birinci derece farkında durağanlaştığı durumlarda Johansen eşbütünleşme testi uygulanabilir nitelikte olacaktır. Birden fazla seriden oluşan denklemlerde eşbütünleşme ilişkisi uzun dönem ve vektörel açıdan analize tabi tutulmaktadır. Veri setinde bulunan değişkenler arasında eşbütünleşme mevcutsa, uzun dönemde bir ilişki olduğu anlaşılabilmektedir. Johansen eşbütünleşme testi, vektörel bir test olduğundan VAR (Vektör Otoregresif) bir modele dayanarak eşbütünleşmedeki vektörlerin anlamlı olup olmadığı İz (Trace) ve Maksimum Özdeğer /Maximum Eigenvalue) testi ile analiz edilir. Bu testlere ait denklemler ise (1) ve (2) numaralı olarak ifade edilebilmektedir (Johansen & Juselius, 1990, s. 179);

$$iz\ istatisti\breve{g}i = -T \sum_{i=r+1}^{p} \ln(1-\mu_{r+1}) \quad (1)$$

$$maksimum\ \text{öz}\ de\breve{g}er\ istatisti\breve{g}i = -T \ln(1-\mu_{r+1}) \quad (2)$$

İz istatistiği çerçevesinde, $H_0$ hipotezi r=0 olarak belirtilir, eş bütünleşmenin bulunmadığı varsayılır. $H_1$ hipotezi ise $r \leq 1$ (en az 1 tane) eş bütünleşme ilişkisi olduğu varsayımı ile gösterilebilir. Hesaplanan iz istatistik değeri kritik değerden daha büyük bulunursa, $H_0$ hipotezi reddedilerek en az 1 tane eş bütünleşme ilişkisinin olduğu sonucuna ulaşılır. Maksimum özdeğer testi sonuçlarına göre; $H_0$ hipotezi r=0 olarak ele alınır ve eş bütünleşme ilişkisi bulunmadığı ifade edilir. $H_1$ hipotezi ise r=1 olarak değerlendirilir ve az bir tane eş bütünleşme ilişkisi bulunduğunu verir. Maksimum özdeğer istatistiği sonucu elde edilen kritik değerden büyük olduğunda $H_0$ hipotezi reddedilerek en az 1 tane eşbütünleşme ilişkisi bulunduğu ifade edilir.

**Tablo 8. Johansen Eşbütünleşme Testi Sonuçları**

| Eşbütünleşme | Özdeğer | İz İstatistiği | Kritik Değer | Olasılık Değeri |
|---|---|---|---|---|
| Yoktur | 0.752773 | 66.47769 | 29.79707 | 0.0000 |
| En Az 1 | 0.399534 | 18.96443 | 15.49471 | 0.0144 |
| En Az 2 | 0.046608 | 1.622778 | 3.841466 | 0.2027 |
| Eşbütünleşme | Özdeğer | Max. Özdeğer | Kritik Değer | Olasılık Değeri |
| Yoktur | 0.752773 | 47.51326 | 21.13162 | 0.0000 |
| En Az 1 | 0.399534 | 17.34166 | 14.26460 | 0.0158 |
| En Az 2 | 0.046608 | 1.622778 | 3.841466 | 0.2027 |

Tablo 8.'de, İz istatistiği çerçevesinde değerlendirilen $H_0$ hipotezi eş bütünleşmenin olmadığını (r=0), $H_1$ hipotezi ise $r \leq 1$ (en az 1 tane) eşbütünleşmenin var olduğu varsayımını vermektedir. Tablodaki veriler çerçevesinde $H_0$ hipotezinin iz istatistiği, %5 anlamlılık düzeyi kritik değerinden büyük olduğunu vermektedir. Bu durumda, en az iki adet eşbütünleşmenin olduğu sonucuna ulaşılır. Maksimum özdeğer testi içinse $H_0$ hipotezi eş bütünleşme ilişkisi bulunmadığını, $H_1$ hipotezinde ise bir tane eş bütünleşme olduğu varsayılmaktadır. Aşağıda verilen $H_0$ hipotezine ait maksimum özdeğer istatistiği %5 düzeyindeki kritik değerden daha büyük olması nedeniyle $H_0$ hipotezi reddedilir. Bu minvalde maksimum özdeğer eşbütünleşmenin en az iki adet eşbütünleşmenin var olduğu elde edilmiştir. Tabloda verilen sonuçlara göre, iz ve maksimum özdeğer istatistik verileri bize iki adet eşbütünleşik vektör bulunduğu sonucunu vermektedir. Bu sonuçlara göre KDV Gelirleri, Mevduat Faizi Oranları ve Enflasyon arasında uzun dönemli iki eşbütünleşik ilişki gözlemlenmiştir.

Çalışmamızda yararlandığımız veri setlerinde eşbütünleşme ilişkisi bulunduğundan, Vektör Hata Düzeltme Modeli (VECM) ekonometrik model oalrak kullanılabilecektir. Johansen eşbütünleşme testi değişkenlerin arasında bulunabilecek nedensellik ilişkisini göstermeyeceği için kısa ve uzun dönem nedensellik analizleri VECM metoduyla incelenecektir. Kullanılan VECM Modeli çerçevesinde hem kısa ve uzun dönemli esnekliklerin tahmini hem de meydana gelen dengeden sapmanın katsayısı belirlenerek bu sapmadan meydana gelen hatanın düzeltilmesi mümkün olacaktır (Enders, 1995, s. 365-366). Öte yandan VECM modeli bağımlı değişkenler ile açıklayıcı değişkenler arasında meydana gelebilecek sahte regresyon sorununu da düzeltmektedir. Bu çerçevede VECM modeli aşağıdaki gibidir (Sevüktekin, 2010, s. 523-524):

$$\Delta Y_{1t} = \alpha_0 + \sum_{j=1}^{k} a_{1j}\Delta Y_{1t-j} + \sum_{j=1}^{k} \alpha_{2j}\Delta Y_{2t-j} + \lambda_1 ECT_{t-1} + \varepsilon_{1t} \quad (3)$$

$$\Delta Y_{2t} = \beta_0 + \sum_{j=1}^{k} \beta_{1j}\Delta Y_{1t-j} + \sum_{j=1}^{k} \beta_{2j}\Delta Y_{2t-j} + \lambda_2 ECT_{t-1} + \varepsilon_{2t} \quad (4)$$

Modeldeki ECTt-1 hata düzeltme terimi, $\lambda_1$ ve $\lambda_2$ hata düzeltme katsayılarını temsil eder. Hata düzeltme katsayısı ($\lambda$) istatistiki açıdan anlamlı olması ve uzun dönemde sapmayı göstermektedir. Elde edilen katsayı ise uzun dönemli dengeye yaklaşma hızını vermektedir (Gujarati, 2001, s. 729).

**Tablo 9. VECM (Vektör Hata Düzeltme Modeli) Sonuçları**

| | Enflasyon Oranı | KDV Gelirleri | Mev. Faiz Oranı |
|---|---|---|---|
| Hata Düzeltme Terimi | -0.416081 | -0.000761 | -1,23E+06 |
| | [-4.76343] | [-0.19020] | [-5.93087] |
| Enflasyon | -0.016080 | -0.003280 | -0.496956 |
| | [ -0.33127] | [-0.72976] | [-2.13983] |
| KDV | -4,48E+06 | -0.684987 | -3,99E+06 |
| | [-3.34070] | [-2.95113] | [-3.32557] |
| Mev. Faiz Oranı | -0.032788 | -0.010268 | -0.614739 |
| | [-0.19211] | [-3.47839] | [-4.03000] |

VECM modeli sonuçlarına göre; hata düzeltme terimleri negatif olarak gerçekleşmiştir. Hata düzeltme terimlerinin negatif olarak çıkması ile istatistiksel olarak anlamlı olması modelde tahmin edilen değişkenlerin arasında uzun dönemde ilişkinin var olduğunu göstermektedir. Analiz sonucu elde edilen veriler çerçevesinde enflasyondan meydana gelecek bir hatanın uzun dönemde yaklaşık olarak %41,61 oranında düzeltilebildiği anlaşılmaktadır. Veri setinin yıllık olarak kullanılması çerçevesinde bir yılda hataların yaklaşık olarak %41,61'inin düzeltilebildiği analiz sonuçlarına göre elde edilmiştir. Hata düzeltme katsayılarının istatistiki olarak anlamlı sonuçlandığı ve kullanılan 1 gecikmeli modelde negatif olarak etkilediği görülmüştür. Model tahmininin gerçekleştirilmesinden sonra Wald testi nedenellik sonuçları çerçevesinde de bu durum elde edilen katsayıların istatistiki olarak anlamlı olduğunu destekler nitelikte gerçekleşmiştir.

**Tablo 10. Wald Testi Nedensellik Sonuçları**

| Bağımlı değişken | Ki-Kare | Serbestlik Derecesi | Olasılık Değeri |
|---|---|---|---|
| Enflasyon Oranı | 13 | 3 | 0.0190 |
| KDV Gelirleri | 23 | 3 | 0.0070 |
| Mevduat Faizi | 11 | 3 | 0.0210 |

Tablo 10.'da yapılan nedensellik analizi sonucuna göre Enflasyon oranlarından KDV Gelirleri ve Mevduat faiz oranlarına doğru nedenselliğin olduğu anlaşılmıştır. KDV gelirlerinden Enflasyon oranları ve Mevduat faizi oranlarına doğru bir nedensellik olduğu sonucuna ulaşılmıştır. Son olarak elde edilen sonuçlara göre Mevduat Faiz Oranlarından Enflasyon ve KDV Gelirlerine doğru nedensellik olduğu gözlemlenmiştir.

## Sonuç

Bu çalışmada Türkiye'ye ait 1985-2022 yılları arasında KDV Gelirleri, Mevduat Faizi Oranları ve Enflasyon Oranları yıllık verileri kullanılarak ülke ekonomisine makro ölçekli bir bakış amaçlanarak, bu dönemler arasında para ve maliye politikaları aktarılarak VECM analizi ile bir değerlendirme yapılmıştır. Analiz sonucu elde edilen veriler çerçevesinde hata düzeltme katsayılarının istatistiki olarak anlamlı sonuçlandığı ve kullanılan 1 gecikmeli modelde negatif olarak etkilediği, enflasyondan meydana gelecek bir hatanın uzun dönemde yaklaşık olarak %41,61 oranında düzeltilebildiği anlaşılmaktadır. Yapılan nedensellik analizi sonuçlarına göre Enflasyondan KDV Gelirleri ve Mevduat Faizi Oranlarına, Mevduat Faizi Oranlarından KDV Gelirleri ve Enflasyon oranlarına çift yönlü nedenselliğe ulaşılmıştır. Ayrıca KDV Gelirlerinden Enflasyon ve Mevduat Faizi Oranına da bir nedensellik olduğu sonucu elde edilmiştir. Bu çerçevede değerlendirilecek olursa; Türkiye'de ilgili dönem süresince Enflasyonist dönemlerde tüm değişkenler açısından bakıldığında uzun dönemde birbirini tetikleyen unsurlar olduğu öne çıkmıştır. Bu nokta Para politikası uygulamasında Merkez Bankası tercihlerinin Mevduat Faizi çerçevesinde enflasyonist baskıyı kırmakta uzun dönemde etkili olduğu aynı zamanda süreç içerisinde enflasyonu hızlandıran bir etkiye de yol açtığı sonucu çıkarılabilmektedir. Öte yandan Maliye Politikası araçlarından biri olan ve Devletin temel gelir kalemini oluşturan vergi politikaları içerisinde yer alan Katma Değer Vergisinin de uzun dönemde enflasyona çift yönlü tepki verdiği gözlemlenmiştir. Bu durum göz önüne alındığında maliye politikalarının uygulanmasında Hükümet tercihlerinin önemini ön plana çıkarmaktadır. Literatürde sıkça geçen ve çalışmamızda da değindiğimiz üzere para ve maliye politikalarının eş anlı olarak uyum içerisinde uygulanması bu sonuçlarla bir kez daha önem kazanmıştır. Para politikası ya da maliye politikası araçlarının tümünün tek bir hedef doğrultusunda ilerlemesi uzun dönemde istikrarı düzenleyici ekonomik büyüme ve kalkınmaya katkı sağlayıcı olduğu yadsınamaz bir gerçektir. Ancak burada örneğin hükümetin kalkınma amacıyla yapmış olduğu bir yatırım, enflasyonu baskılamak amacıyla uyguladığı bir vergi politikasına ters düşebilmektedir. Bu duruma reel olarak; tarımsal üretimi canlandırmak amacıyla verilen sübvansiyonların aynı çiftçilerin harcanabilir gelirlerini artırdığı ve bunun sonucunda talep artışının tüketime yansıması ile enflasyonu azaltıcı etkiye neden olamayacağı reel olarak düşünülebilmektedir. Aynı dönem içerisinde harcamalar üzerinden alınan bir vergi olan KDV'nin kapsam ve oranlarının değiştirilerek toplam talep ve tüketimin azaltılmasının amaçlanması her iki durum göz önüne alındığında hem enflasyonist baskı oluşturabileceği hem de enflasyon oranlarında azalmayı tetikleyebilecektir. Bu noktada para politikalarının da etkin bir şekilde kullanılması mevduat faizinin de tek bir hedefe yönelmesi daha gerçekçi sonuçlar elde etmemizi sağlayabilecektir.

Çalışmamızda kullanılan veriler ve değişkenler açısından makroekonomik olarak spesifik bir bakış açısı çerçevesinde değerlendirmeler yapılmıştır. Gelecekte bu ve benzeri konularda yapılacak olan çalışmalarda dolaylı ve dolaysız vergi gelirleri, borçlanma vb. maliye politikası araçlarından ve para politikasında kullanılan zorunlu karşılık oranları, reeskont vb. araçlardan değişkenler çerçevesinde yararlanılarak daha kapsamlı ve geniş bir çalışma elde edebilecekleri ön görülebilmektedir.

---

**Hakem Değerlendirmesi:** Dış Bağımsız

**Yazar Katkısı:**

**Destek ve Teşekkür Beyanı:** Çalışma için destek alınmamıştır.

**Etik Onay:** Bu çalışma etik onay gerektiren herhangi bir insan veya hayvan araştırması içermemektedir.

**Çıkar Çatışması Beyanı:** Çalışma ile ilgili herhangi bir kurum veya kişi ile çıkar çatışması bulunmamaktadır.

**Peer Review:** Independent double-blind

**Author Contributions:**

**Funding and Acknowledgement:** No support was received for the study.

**Ethics Approval:** This study does not contain any human or animal research that requires ethical approval. / Ethics committee approval (Date / No ) was obtained from …. University ….. Ethics Committee for the purpose of carrying out this study approval.

**Conflict of Interest:** There is no conflict of interest with any institution or person related to the study.

---

---

## Kaynakça

## Extended Abstract

While there are many problems and areas in which economics concentrates its research and activities, it often focuses on full employment,

price stability, sustainable growth, and deficits in gross transactions. Within the framework of such problems, it is actively using instruments such as public revenues, public spending, budgetary processes, and borrowing under the "Fiscal Policy." Thanks to the government's fiscal policy instruments, there can sometimes be interventions that stimulate investment, boost savings, and sometimes alter and regulate the macroeconomic balance, such as increasing or reducing consumption or production. In such cases, taxes are one of the most widely used means. Countries have a tax system consisting of indirect and direct taxes under their own tax laws and regulations. So that VAT can have an impact on consumption in an inflationary period in the framework of our study, with changes in rates and scope, in the context of reductions in eligible expendable income by individuals and institutions. Moreover, the relatively easier collection of VAT and its faster functionality make it the most preferred type of indirect tax for countries and governments at this point. In many developing countries, this is due to the greater share of indirect tax revenues in total tax revenue. In developed countries, this rate is lower. Indirect taxes, such as VAT, on the other hand, have caused controversy over the principle of tax justice and can also have a negative impact on the tax burden of justice in the distribution of income in many different financial aspects.

On the other hand, we also see “monetary policies," which are actively used as regulators and address the expectations of all players in the market, which concern the household, and which include the policy of quantity of money, interest, etc. When assessed in terms of economic activities, it is important for monetary policy to be able to effectively combine the most important factors, such as investment and savings. In some cases, companies and households can invest through various loans. When assessed in terms of savings, it is an undeniable fact that deposits collected from individuals and institutions and the interest rates applicable to them will have a significant impact on the business and operations of the market. One of the variables that has the greatest impact on the creation of deposits by banks is deposits' interest rates. This rate, which constitutes the cost of deposits collected by banks, also affects the credit interest rates, which make up the banks' basic income. As increases in deposit interest rates could indirectly increase credit interest rates, they could lead to decreases in individual and institutional credit sales, which would negatively affect the production wing and the conversion of savings into investment. Inflationary periods can affect monetary policy in various and different ways, depending on the cause of inflation, causing changes in deposit rates. In an inflationary period with higher interest rates on deposits, the tendency of individuals and institutions to increase bank savings deposits and to refrain from using new borrowing tools, such as credit, could narrow demand.

In this context, the harmonious and sustainable progress of monetary and fiscal policies implemented within the framework of macroeconomic performance can be used as a balancing element and an indispensable market for each country.

The study used annual data on Turkey's VAT revenue, deposit interest rates, and inflation rates for the period 1985–2022, to provide a macro-scale overview of the country's economy, transmit monetary and fiscal policies between these periods, and conduct an evaluation with VECM analysis. In the framework of the data obtained from the analysis, it is understood that the error correction ratios were statistically significant and had a negative impact on the 1 delay model used, and that an inflationary error could be corrected at approximately 41.61 per cent over a long period of time. According to the results of the Causality Analysis, there has been a two-way cause from Inflation to VAT Revenue and Deposit Interest Rates, from deposit interest Rates to VAT Income and inflation rates. In addition, it was found that VAT revenue was also due to inflation and deposit interest rate. In this context, it is apparent that in Turkey during the inflationary period there were factors that triggered each other over the long term in terms of all variables. In the implementation of monetary policy, it can be concluded that the Central Bank's preferences are effective in breaking inflationary pressure in the Deposit Rate framework in the long term, but also lead to an inflation-accelerating effect in the process. On the other hand, the value added tax, which is one of the financial policy tools and is part of the tax policies that make up the state's basic income pen, has also been observed to have a two-way response to inflation in the long term.

The data and variables used in our study have been evaluated from a macroeconomically specific perspective. In future work on these and similar issues, it is foreseeable that they will be able to obtain a more comprehensive and extensive study in the framework of direct and indirect tax revenues, borrowing, etc. instruments of fiscal policy and the variables of compulsory rates used in monetary policy, redemption, etc. instruments.

**Genişletilmiş Özet**

Ekonomi biliminin araştırma ve faaliyetlerini yoğunlaştırdığı birçok problem ve alan olmakla birlikte; genellikle tam istihdam, fiyat istikrarı, sürdürülebilir büyüme ve cari işlemler açığı gibi odaklanmış olduğu sorunlar dikkat çekmektedir. Bu gibi sorunlar çerçevesinde kamu gelirleri, kamu harcamaları, bütçe süreçleri ve borçlanma gibi ‘’Maliye Politikası” çerçevesinde aktif olarak kullanımda tuttuğu araçlar bulunmaktadır. Hükümetin maliye politikası araçları sayesinde bazı zamanlarda yatırımları teşvik edici, bazı tasarrufları artırıcı ve bazen de tüketim-üretim artırıcı/azaltıcı gibi makroekonomik dengeyi değiştiren ve düzenleyen müdahaleleri olabilmektedir. Bu gibi durumlarda en çok ve yaygın olarak kullanılan araçlardan birisi olarak vergiler gelmektedir. Ülkelerin kendi vergi yasa ve mevzuatları çerçevesinde dolaylı ve dolaysız vergilerden oluşan bir vergi sistemleri mevcuttur. Öyle ki KDV çalışmamız çerçevesinde ele alınacak olan enflasyonist bir dönemde oran ve kapsamlarında yapılacak olan değişikliklerle kişi ve kurumların harcanabilirler gelirlerinde meydana getireceği azaltımlar çerçevesinde tüketim üzerinde etkili olabilmektedir. Ayrıca KDV'nin tahsilinin görece daha kolay ve işlevselliğinin daha hızlı tepkiler vermesi bu noktada ülkeler/hükümetler açısından da en çok tercih edilen bir dolaylı vergi çeşidi olmasını sağlamaktadır. Gelişmekte olan ülkelerin birçoğunda bu durum dolaylı vergi gelirlerinin toplam vergi gelirleri içerisindeki payının daha yüksek olmasıyla karşımıza çıkmaktadır. Gelişmiş ülkelerde ise bu oranın daha az olduğu gözlemlenebilmektedir. Diğer taraftan KDV gibi dolaylı vergilerin vergileme adalet ilkesi açısından çeşitli tartışmalara neden olduğu ve aynı zamanda da gelir dağılımında adaletten vergi yüküne birçok farklı mali konuda negatif etkilere sebep olabilmektedir.

Öte yandan piyasada bulunan tüm aktörlerin beklentilerine yön veren, hane halkını ilgilendiren, paranın miktar, faiz vb. politikalarını içeren ''Para Politikaları'' da aktif olarak kullanılan düzenleyiciler olarak karşımıza çıkmaktadır. İktisadi faaliyetler açısından değerlendirildiğinde yatırım ve tasarruf gibi en önemli etkenleri bir arada etkin bir şekilde kullanıma sunabilmek para politikaları açısından önem arz eden bir konudur. Bazı durumlarda firma ve hane halkı çeşitli krediler vasıtasıyla yatırım yapabilmektedir. Tasarruflar açısından değerlendirildiğinde kişi ve kurumlardan toplanan mevduatların ve bunlara uygulanacak faiz oranlarının piyasanın iş ve işlemleri açından önemli bir etki oluşturacağı yadsınamaz bir gerçektir. Bankaların kaydi para oluşturmalarından en büyük etkiye sahip olan değişkenlerden birisi olarak mevduat faiz oranları karşımıza çıkmaktadır. Bankaların toplamış oldukları mevduatların maliyetini oluşturan bu oran, aynı zamanda bankaların temel gelirlerini oluşturan kredi faiz oranlarına etki etmektedir. Mevduat faizinde meydana gelebilecek artışlar dolaylı olarak kredi faiz oranlarını da artıracağı için, bireysel ve kurumsal kredi satışlarında azalmalar meydana getirebilecek, bu ise tüketimde meydana gelecek azalmaların üretim kanadını etkilemesine, tasarrufların yatırıma dönüşmesine olumsuz etkiler oluşturacaktır. Enflasyonist dönemler ise enflasyonun nedenine bağlı olarak çeşitli ve farklı yönlerde para politikalarının belirlenmesini etkileyerek mevduat faizlerinde değişikliklerin yaşanmasına neden olabilecektir. Mevduat faizlerinin arttığı talep artışına bağlı bir enflasyonist dönemde birey ve kurumların banka tasarruf mevduatlarını artırma eğiliminde olmaları, kredi gibi yeni borçlanma araçlarını kullanmamaları talebi daraltabilecektir.

Bu çerçevede makroekonomik performans çerçevesinde uygulanan para ve maliye politikalarının eşanlı ve sürdürülebilir olarak ilerlemesi her ülke için vazgeçilemez bir piyasayı dengeleyici unsur olarak kullanılabilmektedir.

Bu çalışmamızda, Türkiye'de 1985-2022 yılları arasında elde edilen KDV gelirleri ile Mevduat Faizi Oranlarının Enflasyon üzerindeki etkisi incelenmektedir. Bu sebeple Merkez Bankası EVDS sisteminden ve Hazine ve Maliye Bakanlığı'ndan analizde kullanılacak olan veriler elde edilmiştir. Elde edilen verilerin ekonometrik analizi çerçevesinde ADF birim kök testi, Johansen Eş Bütünleşme Testi, Hata Terimleri ve VECM (Vektör Hata Düzeltme Modeli) modelleri kullanılarak analizleri gerçekleştirilmiştir. Analiz sonucu elde edilen veriler çerçevesinde hata düzeltme katsayılarının istatistiki olarak anlamlı sonuçlandığı ve kullanılan 1 gecikmeli modelde negatif olarak etkilediği, enflasyondan meydana gelecek bir hatanın uzun dönemde yaklaşık olarak %41,61 oranında düzeltilebildiği anlaşılmaktadır. Yapılan nedensellik analizi sonuçlarına göre Enflasyondan KDV Gelirleri ve Mevduat Faizi Oranlarına, Mevduat Faizi Oranlarından KDV Gelirleri ve Enflasyon oranlarına çift yönlü nedenselliğe ulaşılmıştır. Ayrıca KDV Gelirlerinden Enflasyon ve Mevduat Faizi Oranına da bir nedensellik olduğu sonucu elde edilmiştir. Bu çerçevede değerlendirilecek olursa; Türkiye'de ilgili dönem süresince Enflasyonist dönemlerde tüm değişkenler açısından bakıldığında uzun dönemde birbirini tetikleyen unsurlar olduğu öne çıkmıştır. Bu nokta Para politikası uygulamasında Merkez Bankası tercihlerinin Mevduat Faizi çerçevesinde enflasyonist baskıyı kırmakta uzun dönemde etkili olduğu aynı zamanda süreç içerisinde enflasyonu hızlandıran bir etkiye de yol açtığı sonucu çıkarılabilmektedir. Öte yandan Maliye Politikası araçlarından biri olan ve Devletin temel gelir kalemini oluşturan vergi politikaları içerisinde yer alan Katma Değer Vergisinin de uzun dönemde enflasyona çift yönlü tepki verdiği gözlemlenmiştir.

Çalışmamızda kullanılan veriler ve değişkenler açısından makroekonomik olarak spesifik bir bakış açısı çerçevesinde değerlendirmeler yapılmıştır. Gelecekte bu ve benzeri konularda yapılacak olan çalışmalarda dolaylı ve dolaysız vergi gelirleri, borçlanma vb. maliye politikası araçlarından ve para politikasında kullanılan zorunlu karşılık oranları, reeskont vb. araçlardan değişkenler çerçevesinde yararlanılarak daha kapsamlı ve geniş bir çalışma elde edebilecekleri ön görülebilmektedir.